\documentclass[reprint, amsmath,amssymb, aip,floatfix,superscriptaddress]{revtex4-1}

\usepackage{verbatim}
\usepackage{chapterbib}

\usepackage{xcolor}
\usepackage{graphicx}
\usepackage[caption=false]{subfig}
\usepackage{dcolumn}
\usepackage{bm}
\usepackage{braket}

\usepackage[utf8]{inputenc}
\usepackage[T1]{fontenc}
\usepackage{mathptmx}
\usepackage{etoolbox}

\usepackage{soul}
\usepackage[colorlinks = true, linkcolor = blue, citecolor = blue, urlcolor = blue]{hyperref}
\usepackage{academicons}

\DeclareMathAlphabet{\altmathcal}{OMS}{cmsy}{m}{n}
 
\newcommand{\f}{$\mathcal{F}$}
\newcommand{\orcid}[1]{\href{https://orcid.org/#1}{\textcolor[HTML]{A6CE39}{\aiOrcid}}}

\newcommand{\response}[1]{%
  \bgroup
  \hskip0pt\color{black!80!black}%
  #1%
  \egroup
}

\makeatletter
\def\@email#1#2{%
 \endgroup
 \patchcmd{\titleblock@produce}
  {\frontmatter@RRAPformat}
  {\frontmatter@RRAPformat{\produce@RRAP{*#1\href{mailto:#2}{#2}}}\frontmatter@RRAPformat}
  {}{}
}%
\makeatother

\begin{document}

\preprint{APS/123-QED}

\title{ Coherence-limited digital control of a superconducting qubit using a Josephson pulse generator at 3~K }

\author{M. A. Castellanos-Beltran}
    \affiliation{National Institute of Standards and Technology, Boulder CO 80305, USA}
    \email{manuel.castellanosbeltran@nist.gov}
\author{A. J. Sirois}
    \affiliation{National Institute of Standards and Technology, Boulder CO 80305, USA}
\author{L. Howe}
    \affiliation{National Institute of Standards and Technology, Boulder CO 80305, USA}
     \affiliation{University of Colorado, Boulder CO 80309, USA}
\author{D. Olaya}
    \affiliation{National Institute of Standards and Technology, Boulder CO 80305, USA}
    \affiliation{University of Colorado, Boulder CO 80309, USA}
\author{J. Biesecker}
    \affiliation{National Institute of Standards and Technology, Boulder CO 80305, USA}
\author{S. P. Benz}
    \affiliation{National Institute of Standards and Technology, Boulder CO 80305, USA}
\author{P. F. Hopkins}
    \affiliation{National Institute of Standards and Technology, Boulder CO 80305, USA}

\date{\today}

\begin{abstract}

Compared to traditional semiconductor control electronics (TSCE) located at room temperature, cryogenic single flux quantum (SFQ) electronics can provide qubit measurement and control alternatives that address critical issues related to scalability of cryogenic quantum processors. Single-qubit control and readout have been demonstrated recently using SFQ circuits coupled to superconducting qubits. Experiments where the \response{SFQ electronics} are co-located with the qubit have suffered from excess decoherence and loss due to quasiparticle poisoning of the qubit. A previous experiment by our group showed that moving the control electronics to the 3 K stage of the dilution refrigerator avoided this source of decoherence in a high-coherence 3D transmon geometry. In this paper, we also generate the pulses at the 3 K stage but have optimized the qubit design and control lines for scalable 2D transmon devices. We directly compare the qubit lifetime $T_1$, coherence time $T_2^*$ and gate fidelity when the qubit is controlled by the Josephson pulse generator (JPG) circuit versus the TSCE setup. We find agreement to within the daily fluctuations for $T_1$ and $T_2^*$, and agreement to within  10\%  for randomized benchmarking.  We also performed interleaved randomized benchmarking on individual JPG gates demonstrating an average error per gate of $0.46$\% showing good agreement with what is expected based on the qubit coherence and higher-state leakage. These results are an order of magnitude improvement in gate fidelity over our previous work and demonstrate that a Josephson microwave source operated at 3 K is a promising component for scalable qubit control.

\end{abstract}


\maketitle

%
\begin{figure}[!ht]
    \centering
    \includegraphics[width = .48 \textwidth]{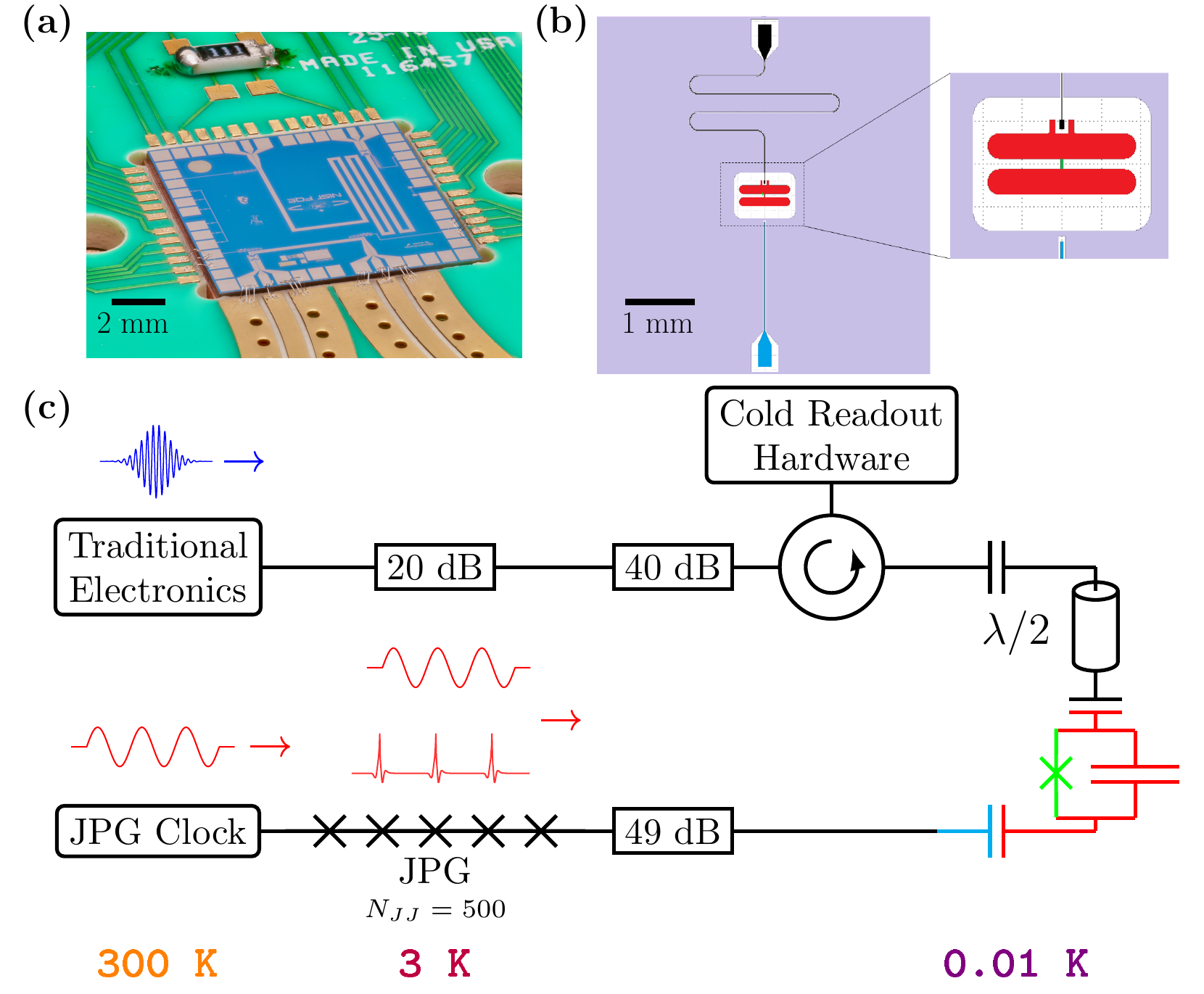}
    \caption{(Color) \response{Digital qubit control using a JPG. \textbf{(a)} Photo of the packaged JPG chip mounted at the 3~K stage of the DR. The drive input (JPG output) is the left (right) coplanar waveguide microwave launch. \textbf{(b)} Layout of the qubit chip mounted on the base temperature stage. The input/output line for the readout cavity is shown in black (top pad) and the direct drive line connected to the JPG to drive the transmon qubit (red) is shown in blue (bottom pad). Inset shows the transmon qubit with the capacitor plates (red) and junction (green).
    \textbf{(c)}  Simplified schematic of the experiment. JPG pulses (along with the larger subharmonic ($\omega_d = \omega_{10} / 2$) sinusoidal drive signal) are routed directly to the qubit drive port.  A commercial TSCE 65~GSa/s arbitrary waveform generator serves as the JPG clock to drive the JPG. The TSCE qubit control/readout synthesizers, and cold readout components are attached to the $\lambda/2$ cavity input port. A detailed description of the setup, including the placement of the attenuation in both lines is shown in the Supplementary Materials.}}
    \label{fig:simplified_schematic}
\end{figure}

 Superconducting quantum circuits are a leading technology for developing quantum computing. However, there are multiple issues with scaling this technology, ranging from refrigeration and power dissipation to the ingress/egress issue related to control, measurement, and error correction of a large number of qubits. Fault-tolerant \response{quantum} computers will require error detection and correction which involves a massive hardware overhead\response{. E}stimates suggest that a general purpose fault-tolerant quantum computer will require millions of physical qubits, far beyond current capabilities.\cite{fowler2012surface, kelly2015state, andersen2020repeated, ai2021exponential}

One possible solution to the ingress/egress issue is to incorporate some of the control electronics at the 3~K stage of the dilution refrigerator (DR). Traditionally, qubit gates and entangling operations are performed using shaped microwave pulses synthesized using traditional semiconductor electronics (TSCE) instrumentation at room temperature. Two options for cryogenic-compatible electronics have recently been proposed: cryo-compatible CMOS (cryoCMOS) and superconductor electronics. Successful integration of DAC/mixers for qubit control \response{at 3~K \cite{vanDijk2020scalable, bardin2019design,ibmcmos} and }at 100~mK \cite{pauka2021cryogenic} using cryoCMOS technology has been demonstrated. Still, significant gaps exist between these devices and a scalable system for qubit control in the areas of gate fidelity, power dissipation, and the accuracy, stability, and repeatability of the control signals.\cite{sirois2020josephson, vanDijk2019impact, ball2016role}  Josephson junction (JJ) based circuits are a promising alternative given the very low \response{($\ll 100~\mu$W)} on-chip power dissipation\cite{mcdermott2018quantum}. 

The use of SFQ pulses has been proposed as a scalable paradigm for digitally controlling qubits,\cite{mcdermott2014accurate, liebermann2016optimal, mcdermott2018quantum,Bastrakova_2022} and was recently demonstrated with SFQ driver and qubit circuits co-fabricated on the same chip.\cite{leonard2019digital} The main limitation of \response{the SFQ circuit (driver) operation proximal} to the qubits is degradation of the qubit lifetimes from quasiparticles created during SFQ pulse generation.\cite{patel2017phonon, martinis2009energy} \response{One solution to suppress this phonon-mediated quasiparticle poisoning of the qubit by the driver is to use a multi-chip module where the SFQ drivers and qubits are fabricated on separate chips.\cite{vincent} Another potential solution, which we pursue in this work, is to move the pulse generation circuitry to the 3~K stage.}
Similar to the work of  Ref. [\onlinecite{leonard2019digital}],  we deliver trains of pulses subresonantly to enact control, giving our device its name of the Josephson pulse generator (JPG). In our previous demonstration of $3$~K JPG control of a high-coherence-time 3-D qubit, we showed no detrimental effects due to quasiparticle poisoning or thermal population on a 3-D qubit. However, a maximum gate fidelity of 98\% was observed. In this report we focus on optimizing a 2-D qubit design, with a dedicated line for control signals, to show nearly coherence-limited single qubit gate operations and demonstrate an order of magnitude improvement in gate fidelity over our previous work.

\begin{figure}
    \centering
    \includegraphics[width = .45 \textwidth]{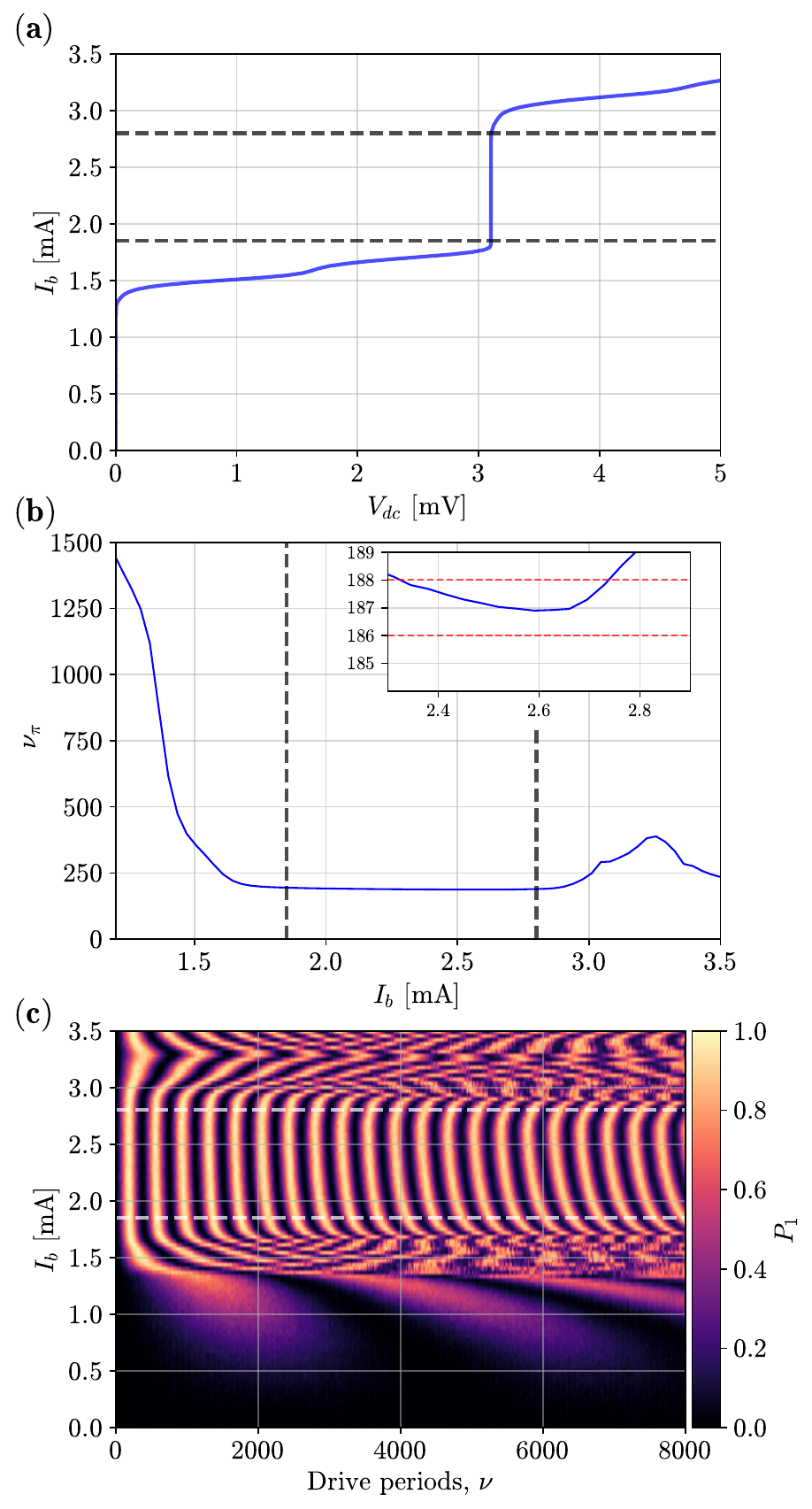}
    \caption{JPG calibration procedure: \textbf{(a)} JPG \mbox{$I$-$V$} curve, shown here with $\omega_d / 2 \pi = 3.0349$~GHz, are first used to establish rough bounds on the range of $I_b$ giving a constant Rabi oscillation period with respect to the number of drive periods $\nu$. From this dc measurement we extract a locking range of 1.8--2.8~mA, which is indicated in all plots with dashed lines. \textbf{(b)} Rabi oscillation scan using the JPG as the drive while scanning current bias $I_b$ and number of drive periods, $\nu$. \textbf{(c)} Extracted number of JPG drive periods  required for a $\pi$ rotation of the qubit $\nu_{\pi}$  versus bias. \response{ The inset shows a zoomed-in area where the Rabi-oscillation period of the qubit is insensitive to variations in $I_b$, with red dashed lines bounding $\nu_\pi$ by $\pm$~1 pulse. For a bias of $I_b = 2.6$~mA we extract $\nu_\pi$ of 187 pulses ($\pi$-gate time of $62$~ns).}} 
    \label{fig:jpg_bringup}
\end{figure}
Due to the Josephson effect, when the phase difference $\delta$ across a JJ evolves by $2\pi$, a voltage pulse {\it V(t)} is generated whose time-integrated area is quantized and equal to the magnetic flux quantum $\Phi_0 \equiv h / 2e$, where $h$ is the Planck constant and $e$ is the electron charge:
\begin{equation}
    \int V dt = \Phi_0.
    \label{eq:quantized_voltage_pulse}
\end{equation}
The duration of this SFQ pulse is approximately \response{given by the characteristic time of the junction defined as} $\tau = \Phi_0 / I_c R_s$,\cite{benz1996pulse} where $I_c$ is its critical current, and $R_s$ is its shunt resistance.\cite{shuntR}

Unlike TSCE microwave qubit drives, we drive the qubit transitions via a train of voltage pulses generated by JJs. If the width of these pulses is much smaller than the qubit transition period, $T_q=2\pi/\omega_{10}$, then during pulse arrival the qubit undergoes a discrete rotation \response{$\delta\theta$ \textit{per pulse} given by}
\begin{equation}\label{eq:delta_theta_sfq}
    \delta \theta = N_{JJ} A C_c \Phi_0 \sqrt{\frac{2 \omega_{10}}{\hbar C_T}},
\end{equation}
where $C_c$ is the control line-qubit coupling capacitance, $C_T$ is the qubit capacitance, $N_{JJ}$ is the number of junctions in our JPG and $A$ is an attenuation parameter that includes intentional as well as parasitic attenuation in the lines.\cite{mcdermott2014accurate} By appropriately designing $N_{JJ}$, $A$ and $C_c$, we achieve appropriate control line thermalization and tip-angle per pulse, $\delta \theta$, with low digitization error.\cite{Logan_PRXQuantum} A train of sharp pulses arriving resonantly to the qubit ($\omega_d = \omega_{10}$), or at a subharmonic ($\omega_d = \omega_{10} / k$, where $k \geq 2$ is an integer), discretely rotates the qubit around the Bloch sphere during pulse arrival. Between pulses, the qubit precesses for $k$ periods at fixed $\theta$, \response{where $\theta$ is the polar angle of the state vector in the Bloch sphere.\cite{mcdermott2014accurate}}

\begin{figure*}
    \centering
    \includegraphics[width = .98 \textwidth]{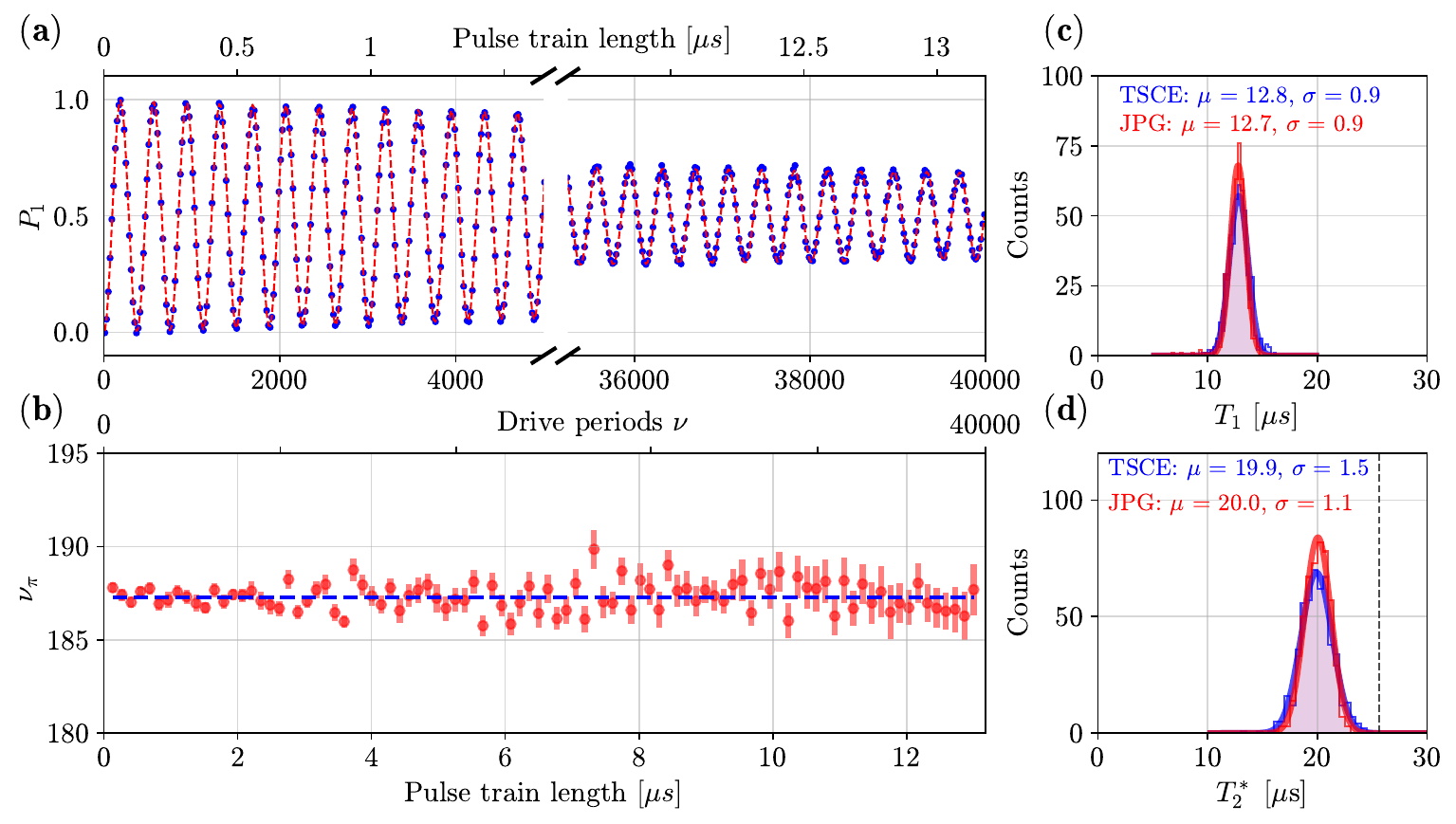}
    \caption{\textbf{(a)} Typical Rabi oscillation driven by the JPG for drive times equal to the longest randomized benchmarking  sequences used in this experiment. From this we extract an approximate decay time of 13.5~$\mu$s, close to the expected decay based on the observed $T_1$ and $T_2^*$.\cite{Andreasdynamic}  \textbf{(b)} Time window analysis of the rabi oscillations shown in (a). The window used is approximately two oscillations. \response{The x-axis in (b) is the same length as in (a) but it has been transformed to time rather than number of drive periods.} We do not observe significant deviation in the number of pulses for a $\pi$ rotation $\nu_\pi$ across the entire experiment. \textbf{(c)} Comparison of the measured qubit lifetime $T_1$ and \textbf{(d)} Ramsey coherence time $T_2^*$ using a TSCE setup and a JPG at 3~K. Histograms and Gaussian fits of both the  $T_1$ and $T_2^*$ distributions show excellent agreement in the mean ($\mu$) and standard deviation ($\sigma$). \response{The black dashed line represents the $2T_1$ limit for $T_2$.}}
    \label{fig:T1_T2_comparison}
\end{figure*}


Since the voltage pulses are generated at the 3~K stage and routed to the qubit using coaxial cabling, the amplitude of a single SFQ pulse is too small to efficiently drive the qubit through the necessary drive line attenuation between the 3~K and the mixing chamber stage. Thus, an amplified multi-SFQ pulse is needed; this is possible using long series arrays of $N_{JJ}$ junctions whose pulse amplitudes add coherently \cite{christine} in topologies similar to those used in superconducting primary voltage standards.\cite{rufenacht2018impact} This multi-SFQ pulse has area $N_{JJ} \Phi_0$. Arrays with $N_{JJ} \sim 10^2-10^4$ are required if located at 3~K, with $N_{JJ}$ depending on $C_c$ and $A$. 
We estimate $C_c\simeq 0.3$~fF based on EM simulations and assume $A=49$~dB, the amount of explicit attenuation placed in the qubit drive line. The JPG has $N_{JJ} = 500$, $I_c = 3.05$~mA, and $R_s = 6.93$~m$\Omega$, resulting in a JJ characteristic frequency of $f_c = 1 / \tau = 10.2$~GHz. \mbox{Fig.~\ref{fig:simplified_schematic}(a)--(c)} show an image of the packaged device, \response{a diagram of the qubit chip,} and a schematic of the experiment\response{, respectively}. $N_{JJ}$ was specifically chosen \response{to obtain a $\delta\theta\simeq\altmathcal{O}(1)$~degree; based on the measured and simulated parameters of the device, we expect $\delta\theta=2.6^\circ$ from Eq. 2. Later we show a $\pi$ rotation corresponds to 187 pulses, i.e., $\delta\theta\simeq 0.96^\circ$; the difference can be explained when we include the finite width of the pulses\cite{Logan_PRXQuantum} as well as the 1-2~dB additional attenuation due to the coaxial cable used to route the pulses from 3~K to the qubit.} The characteristic frequency  $f_c$ was chosen high enough so that the pulse width was much smaller than the qubit period but low enough to ensure ample operating margins at the reduced drive frequency $f_d/f_c\simeq 0.3.$\cite{benz1996pulse}

\response{The JPG is driven using a sinusoidal signal at $k\geq2$ because there is no isolation between the drive input and device output (see Fig. \ref{fig:simplified_schematic}c). Otherwise, if k=1 was used, the large input signal used to drive the JPG (the clock signal) would dominate the qubit control dynamics.} Generation of an integer number of JPG pulses $\ell$ is performed by sending an integer number of sinusoidal drive periods, $\nu$. Under certain bias parameter settings, called the locking range, there is a one-to-one correspondence between the number of JPG pulses generated and the number of sinusoidal drive periods ($\nu = \ell$). Orthogonal axis control is realized by phasing the drive signal relative to a timing reference. More details are found in Ref. [\onlinecite{Logan_PRXQuantum}].


We use a standard 2D qubit design with a $\lambda/2$~transmission line resonator capacitively coupled to a transmon qubit (Fig \ref{fig:simplified_schematic}b). Unlike our previous work,\cite{Logan_PRXQuantum} the qubit has a dedicated drive line to avoid cavity-induced distortions of the JPG signal. This direct coupling allows for a significant reduction in the JJ array size $N_{JJ}$.  
The TSCE control and readout line are both connected to the port of the resonator (0.85~MHz coupling rate). With this setup, a direct comparison of qubit performance with both control schemes is possible during the same cooldown. Qubit readout is performed by probing the qubit-state-dependent frequency shift of the cavity. A Josephson parametric amplifier\cite{castellanos2007widely} (JPA) is operated with a phase-insensitive gain of 15~dB to enable single-shot measurements.
The same readout procedure and instrumentation is used for both TSCE and JPG measurements.


To summarize the calibration method explained in Ref.~[\onlinecite{Logan_PRXQuantum}], we first characterize the qubit with the TSCE setup and then determine the JPG operating parameters. Specifically, the RF JPG clock power and dc current bias $I_b$ locking ranges are determined. This is shown in in Fig. \ref{fig:jpg_bringup}. When applying the appropriate RF drive at a frequency $f_d$, a constant voltage Shapiro step\cite{vanduzer} appears at
\begin{equation}
    V = N_{JJ} \Phi_0 f_d.
    \label{eq:shapiro_voltage}
\end{equation}
For any $I_b$ on the Shapiro step the device is locked. Thus, we first maximize the locking range by determining the drive power giving the largest Shapiro steps. Fig.~\ref{fig:jpg_bringup}(a) shows an example JPG \mbox{$I$-$V$} curve with locking range defined by black dashed lines. As we are restricted to subharmonic drive frequencies, $k = 2$ maximizes the locking range by making $f_d$ as close as possible to $f_c$ \cite{benz1996pulse} \response{and provides the highest fidelity (fastest) gates by reducing gate infidelity due to qubit decoherence.}

Once an appropriate range of bias and RF power has been established, we measure JPG-induced Rabi oscillations to characterize the JPG-qubit coupling. At the optimal drive power, we measure the number of JPG drive periods $\nu$ required for a $\pi$ rotation of the qubit, $\nu_\pi$, versus $I_b$. Results of this procedure are shown in Fig.~\ref{fig:jpg_bringup}(b)--(c). By fitting these Rabi oscillations at constant $I_b$, we obtain the number of pulses for a $\pi$ rotation $\nu_\pi(I_b)$ and we look for regions where $\nu_\pi$ is insensitive to the number of Rabi periods (i.e. drive time). This demonstrates locking of the JPG. As discussed in Ref. [\onlinecite{Logan_PRXQuantum}], there is a small variation of $\nu_\pi$ due to a bias-dependent pulse width and since the characteristic time of our junctions is $\Phi_0/(I_cR_s) \simeq 90-100$~ps, the JPG pulses cannot be approximated as simple delta functions. 


\begin{figure*}
    \centering
    \includegraphics[width = .98 \textwidth]{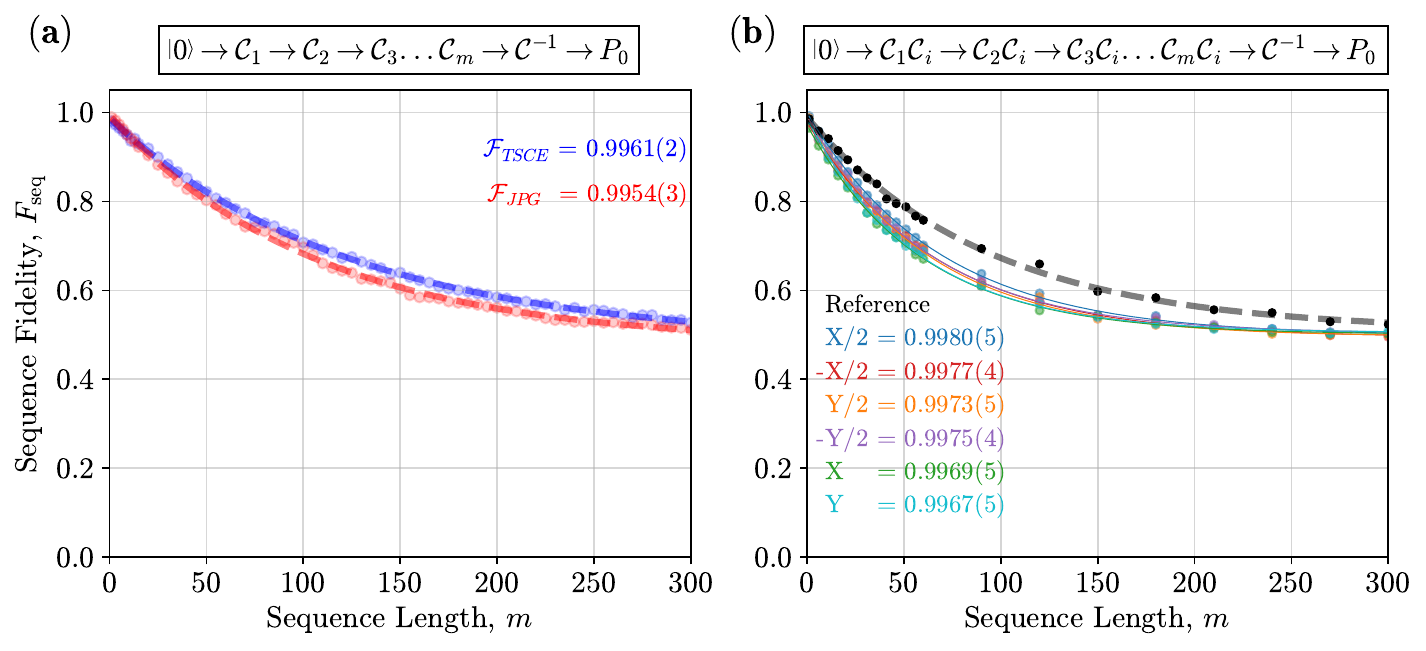}
    \caption{(Color) \textbf{(a)} Depolarizing curve for single qubit RB using  the full Clifford set. Both TSCE and 3~K JPG qubit control setups have very similar performance. Solid lines are a fit to Eq.~\ref{eq:rb_fidelity}. We extract an average error per gate of $r_{TSCE} = (3.9 \pm 0.2) \times 10^{-3}$ and $r_{JPG} = (4.6 \pm 0.3) \times 10^{-3}$, showing an improvement of almost an order of magnitude compared to previous JPG implementations. Uncertainties in $r=1-\altmathcal{F}$ are determined as the standard error of the fits to Eq.~\ref{eq:rb_fidelity}. \textbf{(b)} Depolarizing curves for the six interleaved RB gate sequences and \response{for the shorter reference sequence (without $\altmathcal{C}_i$) in black} using the JPG for the drive. Inset list shows the measured fidelities.
    }
    \label{fig:jpg_rb}
\end{figure*}


 One of the issues observed in Ref. [\onlinecite{leonard2019digital}] was a reduction of the qubit lifetime and coherence as more pulses were generated in the SFQ driver. Here we show that the JPG pulse drive does not reduce the qubit coherence over the long timescales used for randomized benchmarking (RB) characterization by performing long Rabi stability tests as shown in Fig. \ref{fig:T1_T2_comparison}a. The observed relaxation in the Rabi oscillations is consistent with the observed energy relaxation ($T_1$) and decoherence ($T^*_2 $) measured using TSCE. \response{We also characterized the stability of the locking range by analyzing the stability of the Rabi frequency during an extended Rabi experiment. Shown in Fig. \ref{fig:T1_T2_comparison}(b) is the Rabi frequency, extracted from Fig. \ref{fig:T1_T2_comparison}(a) using a time window analysis, as the Rabi experiment progresses in time. The time window used is approximately two oscillations.}

 Once the appropriate $\nu_\pi$ has been established, \response{we perform a side-by-side comparison of the TSCE and JPG setups through measurements of $T_1$, $T_2^*$ and average gate fidelity  $\altmathcal{F}$.} For the $T_1$ comparison, a JPG $\pi$ rotation \response{($X_{\pi}$)} is constructed of $\nu_\pi = 187$ drive periods (62~ns drive time). For the $T^*_2$ comparison a JPG $X_{\pi/2}$ rotation is created with a $\nu_\pi /2 = 94$ period drive waveform. \response{Histograms of the extracted $T_1$ and $T_2^*$ values from 500 measurements each are compiled in Fig.~\ref{fig:T1_T2_comparison}c and d, respectively.} Excellent agreement between the two setups is found in the measured values for $T_1$ and $T_2^*$, showing that JPG operation does not increase relaxation or de-phasing from quasiparticle poisoning.

After demonstrating basic qubit control with JPG pulse trains, we characterize the fidelity of SFQ-based gates. We perform RB for the full single-qubit Clifford set: we apply a sequence of $m$ random gates ($\altmathcal{C}_m$) for the base RB sequence (Fig. \ref{fig:jpg_rb}a), or $m$ random gates with the gate under test ($\altmathcal{C}_i$) interleaved in the sequence for the interleaved randomized benchmarking (IRB) sequence (Fig. \ref{fig:jpg_rb}b). This is followed by a single gate ($\altmathcal{C}^{-1}$) to return the qubit to the $\ket{0}$ state. Following the above procedure, one can plot the average post-sequence fidelity, defined as the average probability of returning the qubit to $\ket{0}$ after each sequence, and fit the two resulting depolarizing curves (with and without the interleaved gate $\altmathcal{C}_i$) to a power law.
\begin{equation}
    F_{\mathrm{seq}}(m) = a p^m + b,
    \label{eq:rb_fidelity}
\end{equation}
where the constants $a$ and $b$ encapsulate state preparation and measurement (SPAM) errors and errors on the final gate. For single qubit gates, the depolarizing parameter, $p$, is related to the per-gate error, $r$, by
\begin{equation}
    r = \frac{1}{2}(1 - p).
    \label{eq:per_gate_error_rb}
\end{equation}
Or alternatively, gate fidelity $\altmathcal{F}=1-r$.

TSCE gates consist of traditional microwave pulses at the qubit frequency shaped using a Gaussian with a width of $\sigma_{\mathrm{TSCE}} = 15$~ns, and truncated at $\pm 2 \sigma_{\mathrm{TSCE}} $ for a total time length designed to closely match the JPG $\pi$-gate time. In both cases, we include a 5~ns buffer idle time around each gate. In Fig.~\ref{fig:jpg_rb}a we show example depolarizing curves for RB for both the JPG and TSCE. Sequence fidelity is defined as the probability of return to the qubit ground state following the final pulse of the RB sequence. From the fits we obtain $r_{TSCE} = (3.9 \pm 0.2) \times 10^{-3}$ and $r_{JPG} = (4.6 \pm 0.3) \times 10^{-3}$, where the uncertainties are from the error obtained from the fit to Eq.~\ref{eq:rb_fidelity}. The observed gate error in both cases is within 50--70\% of the \response{calculated  gate error due solely to qubit dissipation and decoherence}:\cite{rol2017restless} $3.0 \times 10^{-3}$ for TSCE and $2.7 \times 10^{-3}$ for the JPG. \cite{fidtime}

We then performed IRB on several individual gates.\cite{magesan2012efficient, chen2018metrology, rol2017restless} Fig.~\ref{fig:jpg_rb}b shows examples of depolarizing curves for a gate subset  ($\pm X_{\pi/2}, X_\pi, \pm Y_{\pi/2}$ and $Y_\pi$) when driven by the JPG. These individual measurements are consistent with the observed RB error of $r_{JPG} = 4.6 \times 10^{-3}$. Although we have not performed measurements that fully account for the sources of the gate errors, these IRB results show the JPG-based control setup operates nearly at the qubit coherence-limit. 

\response{Given the sub-resonant drive of the qubit, one likely source of errors is leakage to the second excited state ($\ket{f}$).\cite{mcdermott2014accurate} We have performed simulations\cite{johansson2012qutip} of this leakage, characterized by the pure-state density matrix occupation $\rho_{ff}$, assuming the measured parameters of our system (anharmonicity, pulse width, and JPG-qubit coupling strength); the results are shown in Fig.  \ref{fig:simulation} (see Supplementary Material for a more detailed explanation of the simulations).} Based on the measured $I_cR_s$ value, the minimum width of the JPG pulses delivered to the qubit is $\sigma_{JPG}=17$~ps, when parameterized by a Gaussian pulse. Room temperature measurements of the pulses place an upper bound on the width of $\sigma_{JPG}=35$~ps.\cite{Logan_PRXQuantum} Together, these two estimates correspond to a pulse width range of \mbox{$\sigma_n = $0.1--0.2} in units normalized by the qubit period $T_q$ ($\sigma_n=\sigma_{JPG}/T_q$). From Fig.~\ref{fig:simulation}, we expect leakage of $\rho_{ff} \simeq (0.7 - 2) \times 10^{-3}$ for an $X_{\pi}$ gate (leakage is approximately constant below $\sigma_n = 0.1$ and decreases as $\sigma_n$ increases past this threshold). The magnitude of this error is comparable to the observed difference between the measured gate fidelity and the calculated coherence-limited one.

\begin{figure}[t!]
    \centering
    \includegraphics[width = .48 \textwidth]{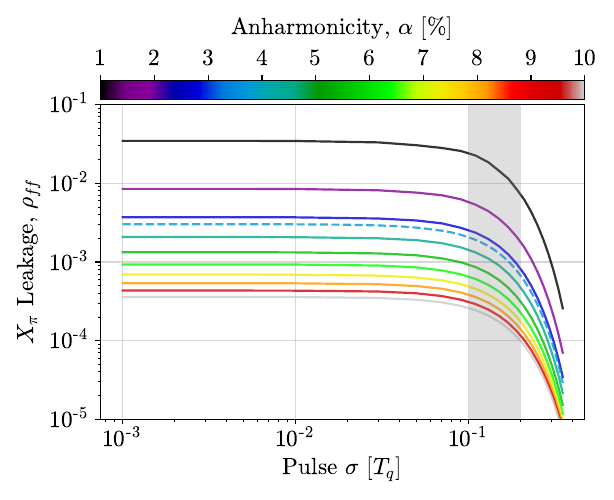}
    \caption{(Color)  Leakage to the second excited state as a function of $\sigma_n$ for different anharmonicities. For $\pi$ rotations of transmons with typical anharmonicities of 3--7\%, a minimum pure-state leakage of approximately 0.07--0.3\% is observed \response{in the $\delta$-function pulse-width limit and driving at the $k=2$ subharmonic. For our case, with $\alpha = 3.3$\% (dashed curve) and $\sigma_n=0.1-0.2$ (vertical gray band), leakage of $\rho_{ff}=0.07-0.2$~\% can be expected and would explain the small disagreement between the observed gate fidelity and the calculated coherence-limit.}
    }

    \label{fig:simulation}
\end{figure}


In conclusion, by optimizing the JPG-qubit coupling, we have experimentally demonstrated nearly coherence-limited digital control of a qubit using JPG pulses generated at 3~K. The observed gate fidelity of 99.54\% using RB agrees with the measured individual gate fidelities using IRB, and are consistent with the limits imposed by qubit relaxation together with higher state leakage due to sub-resonant drive of the qubit. Further improvement in gate fidelities could be achieved by using qubits with longer coherence times, using shorter gates, using higher anharmonicity qubits (since this reduces leakage) and/or utilizing pulse generators operating at frequencies above the qubit frequency and capable of variable pulse-delivery timing.\cite{liebermann2016optimal,McDermott_shortpulses}

\section*{Supplementary Material}
See  supplementary material for specifications of the device under test, a description of the simulations, and a detailed schematic of the measurement setup.

\begin{acknowledgements}
    We acknowledge the NIST Advanced Microwave Photonics Group for the JPA.  We thank Robert McDermott, Britton Plourde, Chuan-Hong Liu and Maxim Vavilov for fruitful discussions.
\end{acknowledgements}

\section*{Data Availability Statement}

The data that support the findings will be available in NIST's Management of Institutional Data Assets (MIDAS) at https://doi.org/10.18434/mds2-2932 following an embargo from the date of publication to allow for commercialization of research findings.

\section*{Author Declarations}

\subsection*{Conflict of Interest}

The authors have no conflicts to disclose.

\subsection*{Author Declarations}

\textbf{M. A. Castellanos-Beltran:} Conceptualization (equal); Resources (equal); Formal analysis (lead); Data curation (Lead); Visualization (Lead); Writing draft (lead); Writing – review \& editing (equal). \textbf{A. J. Sirois}: Conceptualization (equal); Resources (equal); Formal analysis (supporting); Writing – review \& editing (equal). \textbf{L. Howe}:  Data curation (equal); Resources (equal); Formal analysis (supporting); Visualization (equal); Writing – review \& editing (equal). \textbf{D. Olaya}:  Resources (lead). \textbf{J. Biesecker}:  Resources (supporting). \textbf{S. P. Benz}: Project administration (lead); Supervision (equal); Writing –review \& editing (equal). \textbf{P. F. Hopkins}: Project administration (lead); Supervision (equal); Writing – review \& editing (equal).

\bibliography{references}



\clearpage
\clearpage
\pagenumbering{arabic}
\setcounter{page}{1}

\let\oldbibliography\bibliography
\renewcommand{\bibliography}[1]{}

\title{ Supplementary Material: Coherence-limited digital control of a superconducting qubit using a Josephson pulse generator at 3~K }

\maketitle
\onecolumngrid
\section{Device Details}

\response{Details of the qubit and cavity system are shown in the table below:}
\renewcommand{\thetable}{S1}
\begin{table}[ht]
\caption{}
\begin{center}
\begin{tabular}{| c | c |}
\hline
Cavity frequency $\omega_c $ & $2\pi \times $7.4498 GHz  \\
\hline
Cavity linewidth $\gamma_c$ & $2\pi \times $ 0.850 MHz  \\
\hline
Qubit frequency $\omega_{10}$ &  $2\pi \times $ 6.0698 GHz  \\
\hline
Qubit anharmonicity $\alpha$   &  214 MHz  \\
\hline
Dispersive Shift  $2\chi$ &  $2\pi \times $  920 kHz \\
\hline
\end{tabular}
\end{center}
\end{table}

\renewcommand{\thefigure}{S1}
\section{Simulations}
\begin{figure*}[ht!]
    \centering
    \includegraphics[width = .48 \textwidth]{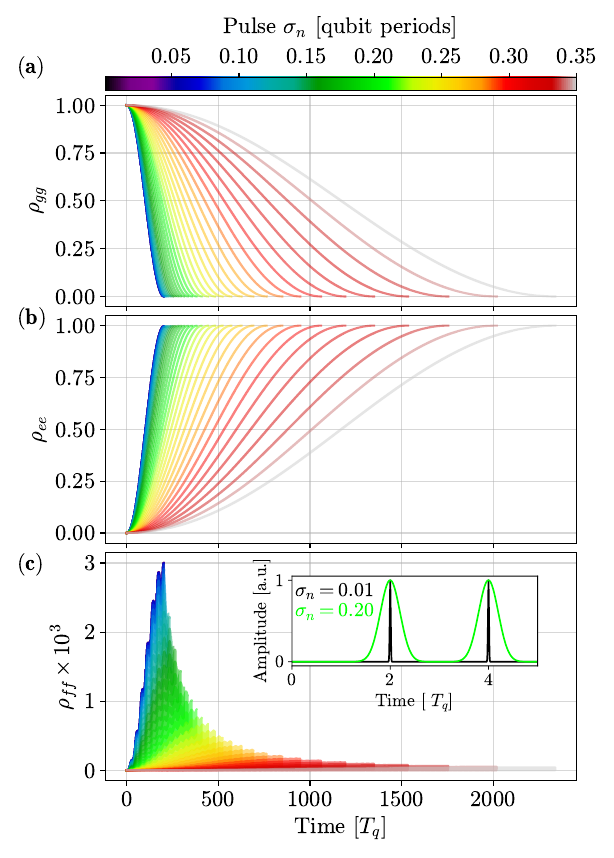}
    \caption{(Color) \response{Simulation of digital control ($X_\pi$ rotation) of a three-level transmon system using a train of Gaussian pulses while omitting dissipation. Pulses are delivered at subharmonic $k = 2$: $\omega_d = \omega_{10} / 2$. \textbf{(a),~(b),~(c)} The diagonal elements of the pure-state density matrix $\rho$ of the ground  ($g$), first ($e$) and second ($f$) excited states, respectively, during delivery of the Rabi pulse train. As the pulse width $\sigma_n$ increases, the number of digital pulses required for a $\pi$ rotation also increases (see Ref. [\onlinecite{Logan_PRXQuantum}]). We used an anharmonicity of 3.3\% to match our physical qubit. The leakage to the second excited state is represented by $\rho_{ff}$ and depends strongly on the normalized pulse width $\sigma_n$ in the range $\sigma_n=0.1-0.2$ we are operating in. Inset: Two extreme examples of the Gaussian pulses used in our simulation, having a normalized width $\sigma_n$ of 0.01 and 0.2 (an estimate of the widest possible pulses in our experiment). Pulses are delivered at $\omega_d = \omega_{10} / 2$, so a pulse arrives every other qubit period.} }
    \label{fig:simulation_appendixa}
\end{figure*}

\response{
In Ref. [\onlinecite{Logan_PRXQuantum}] we performed simulations that included finite pulse width effects on the qubit rotation from multi-SFQ pulses generated by the JPG. To generate Fig. 5 (main manuscript) and Fig. 6 (below), we also included the effect of anharmonicity of the qubit on the leakage to a higher state.  
For this, the main modification compared to  Ref. [\onlinecite{Logan_PRXQuantum}] is that we have modeled the qubit as a three-level system following the description in Ref. [\onlinecite{mcdermott2014accurate}] for the evolution of the three-level system case. We simulate these effects \cite{johansson2012qutip} using a train of Gaussian pulses delivered at $\omega_d = \omega_{10} / 2$ to evolve the qubit state and study the fidelity of $X_\pi$ rotations as a function of the normalized pulse width $\sigma_n$.  With the modification of the extra state in the qubit, the Hamiltonian for the system is now 

\renewcommand{\theequation}{S1}
\begin{equation}
    \hat{H} = \hat{H}_0 + \hat{H}_d =\hbar \hat{\Sigma}_z + i\hbar s_d(t) \hat{\Sigma}_y
    \label{eq:sim_hamiltonian}
\end{equation}
where
\renewcommand{\theequation}{S2}
\begin{equation}
    \Sigma_z=\begin{pmatrix} 0 & 0 & 0\\ 0 & \omega_{10} & 0\\ 0 & 0 & \omega_{10}+\omega_{21}\end{pmatrix}, 
    \Sigma_y=\begin{pmatrix} 0 & -\Omega_d & 0\\ \Omega_d & 0 & -\sqrt{2}\Omega_d\\ 0 & \sqrt{2} \Omega_d& 0\end{pmatrix}.
\end{equation}
$\Omega_d$ and $s(t)$ describe the coupling strength and the pulse train time-dependent amplitude, respectively. In our case, this means that 
\renewcommand{\theequation}{S3}
\begin{equation}
\Omega_d=N_{JJ} A C_c \sqrt{\frac{\omega_{10}}{2 \hbar C_T}}
\end{equation}
and $s(t)$ consists of Gaussian pulses as shown in Fig. \ref{fig:simulation_appendixa}(c) (inset) with a voltage amplitude such that their integral is quantized and given by $\Phi_0$. Results of the simulations are shown using the values of the diagonal matrix representing the density matrix $\rho$. We use the probability ($\rho_{ee}=P_1$) of the 3-level system of being in the first excited state $\ket{e}$ to calculate the value of $\nu_\pi(\sigma_n)$, where $\nu_\pi(\sigma_n)$ is the number of pulses which maximize $\rho_{ee}$ as a function of the Gaussian pulse width $\sigma_n=\sigma/T_q$. $T_q=2\pi/\omega_{10}$ is the qubit period. As in the pure two-level case  simulated in Ref. [\onlinecite{Logan_PRXQuantum}], when $\sigma_n$ is increased, $\delta \theta$ diminishes and more pulses are required to realize a $\pi$ rotation. We must in turn recalculate $\nu_\pi(\sigma_n)$ for each pulse width. Since the goal of these simulations is to consider leakage to the second excited state $\ket{f}$, we do not include loss. Figures \ref{fig:simulation_appendixa}(a-c) show the results of the simulations, where we plot the probability of occupancy for the ground, first and second excited states as a function of normalized time in units of $T_q$. In Fig. \ref{fig:simulation_appendixb}, we show the expected leakage for different anharmonicities in the small pulse-width limit and compare it with theory presented in Ref. [\onlinecite{mcdermott2014accurate}].

}

\renewcommand{\thefigure}{S2}
\begin{figure*}[ht!]
    \centering
    \includegraphics[width = .98 \textwidth]{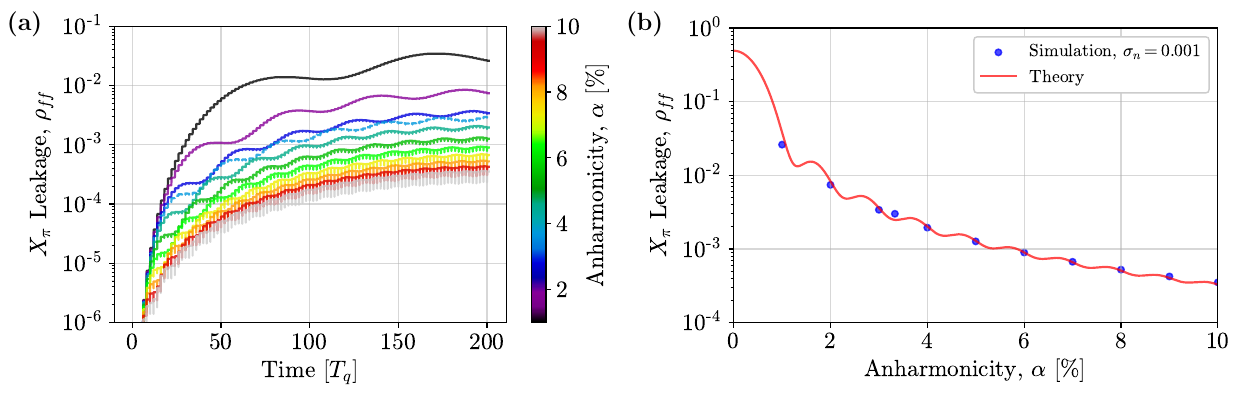}
    \caption{(Color) \textbf{(a)} Simulated leakage to the second excited state $\ket{f}$ in the limit of small pulse width ($\delta$-function) for different anharmonicities as a function of time during  an interval equivalent to a $\pi$ rotation. \textbf{(b)} Expected leakage to the $\ket{f}$~state as a function of anharmonicity after a $\pi$~ pulse. \response{Red curve is the approximation described in Ref.[\onlinecite{mcdermott2014accurate}] adjusted for a k=2 subharmonic drive. Blue dots are numerical simulations.}}
    \label{fig:simulation_appendixb}
\end{figure*}

\newpage

\section{Schematic}

\response{A detailed diagram for the experimental setup is shown in Fig. \ref{fig:schematic}.} 

\renewcommand{\thefigure}{S3}
\begin{figure*}[ht!]
    \centering
    \includegraphics[width = .98 \textwidth]{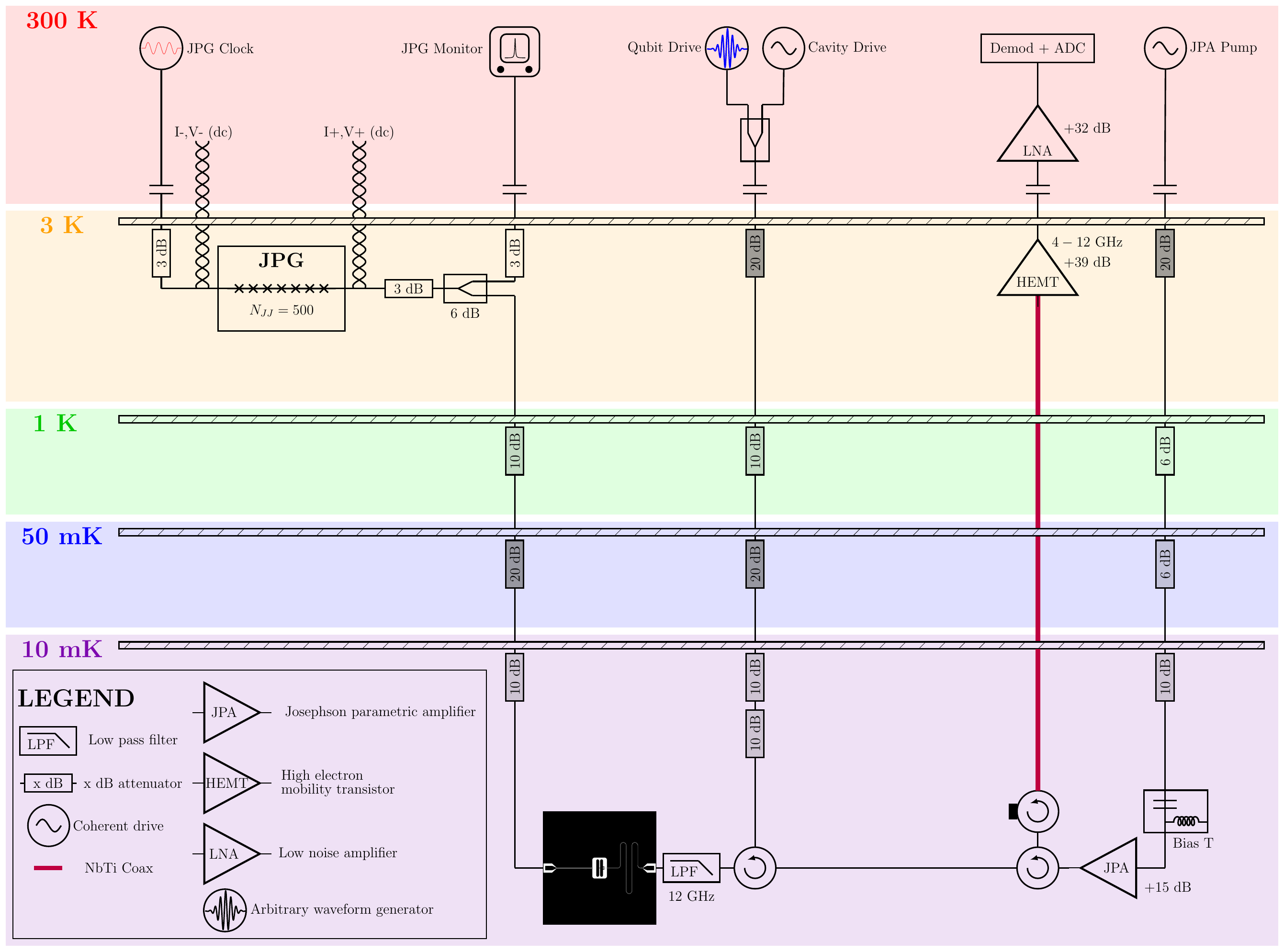}
    \caption{(Color) Detailed schematic of the  the circuitry employed in the cryostat for qubit measurement and control experiments. \response{For comparison, we have used the JPG Monitor line as a way to drive the qubit using TSCE and we have not seen any difference in gate fidelity using TSCE when driving using this line or the cavity drive line.}}
    \label{fig:schematic}
\end{figure*}




\end{document}